\pdfoutput=1
\documentclass[conference]{IEEEtran}
\usepackage{fancyhdr}
\usepackage{cite}
\usepackage{amsmath,amssymb,amsfonts}
\usepackage{graphicx}
\usepackage{booktabs}
\usepackage{xcolor}
\usepackage{url}

\begin{document}

\title{CQTwin: A Digital-Twin Toolkit for Operational Analytics of HPC Schedulers}

\author{
  \IEEEauthorblockN{Kanglin Xu\IEEEauthorrefmark{1},
                    Yash Kurkure\IEEEauthorrefmark{1},
                    Yihe Zhang\IEEEauthorrefmark{1},
                    Michael E.\ Papka\IEEEauthorrefmark{2}\IEEEauthorrefmark{1},
                    Zhiling Lan\IEEEauthorrefmark{1}\IEEEauthorrefmark{2}}
  \IEEEauthorblockA{%
    \IEEEauthorrefmark{1}Department of Computer Science,
    University of Illinois Chicago, Chicago, IL, USA\\
    \IEEEauthorrefmark{2}Argonne National Laboratory, Lemont, IL, USA}
}

\maketitle

\thispagestyle{fancy}
\fancyhf{}
\lfoot{\parbox[t]{\textwidth}{\scriptsize
Accepted at the 1st International Workshop on HPC Operational Data Analytics (HPC-ODA 2026),
held with SC26, November 15--20, 2026, Chicago, Illinois, USA.\\
\copyright~2026 IEEE. Personal use of this material is permitted. Permission from IEEE must be
obtained for all other uses, in any current or future media, including reprinting/republishing
this material for advertising or promotional purposes, creating new collective works, for resale
or redistribution to servers or lists, or reuse of any copyrighted component of this work in
other works.}}
\renewcommand{\headrulewidth}{0pt}
\renewcommand{\footrulewidth}{0pt}

\begin{abstract}
HPC facilities collect years of job scheduler records, yet much of this operational data remains underused because accessible tools for turning it into actionable insight are scarce.
We present \textbf{CQTwin}, an open-source digital-twin toolkit for operational analytics of HPC scheduling data. CQTwin addresses this gap through two complementary modes: offline trace replay and online event-driven simulation.
Built on the Python-based CQSim discrete-event simulator, CQTwin combines a fidelity-validated, high-performance C++ engine with an interactive browser-based dashboard and a Snapshot Lab, enabling facility staff, operators, and researchers to characterize workloads, compare scheduling policies, and explore what-if scenarios.
As an offline twin, CQTwin replays archived workload traces under alternative policies to show how the same workload would have unfolded under different scheduling decisions.
As an online twin, it consumes job events as they arrive and projects how alternative policies would process the currently queued jobs, providing guidance for real-time decision-making.
The dashboard supports the complete analysis workflow, including workload characterization, experiment configuration and monitoring, and multidimensional comparison of simulation results, allowing users to determine visually which policies best suit a workload and what trade-offs they entail.
We validate CQTwin using production traces from the Argonne Leadership Computing Facility spanning 2016--2023.

\end{abstract}


\section{Introduction}
\label{sec:intro}

High-performance computing (HPC) facilities continuously generate rich streams of operational data, including system logs, hardware telemetry, and job logs. Among these sources, scheduler records provide a particularly comprehensive view of facility operation: the lifecycle of every submitted job, how users request resources, how workloads evolve over time, and how scheduling decisions affect wait time, utilization, and fairness~\cite{amvrosiadis2018diversity}.
Public archives of production workload traces have therefore become a foundation for reproducible workload characterization and scheduling research~\cite{feitelson1996,feitelson2014pwa}.

A persistent gap, however, separates \emph{collecting} scheduler data from \emph{acting on} it. Operations teams may accumulate years of scheduler telemetry, yet lack accessible tools for translating those records into answers to practical policy questions.
These questions generally require two distinct modes of analysis.
Some are retrospective: \emph{What would the average wait time have been under a different priority policy? Would another backfilling strategy improve fairness at the cost of utilization?} These are \textbf{offline} questions: the full trace is known, a candidate policy can be replayed from beginning to end, and their aggregate outcomes can be compared.
Other questions arise during operation: \emph{Given the jobs queued and running now, which policy is likely to clear the backlog fastest? Should the policy change as the workload shifts?} These are \textbf{online} questions: jobs become known only when they arrive, and decisions must be evaluated without access to future submissions.

A digital twin of the batch scheduler can support both forms of analysis.
Offline, the twin can replay historical workloads under counterfactual policies without disrupting production.
Online, it can mirror the scheduler-visible state and project the near-term consequences of alternative decisions before they are committed. Together, these modes can turn scheduler records from passive archives into an operational decision-support resource.

Turning scheduler data into operational insight, however, remains harder than it should be.
First, existing open-source trace simulators~\cite{klusacek2010alea,simakov2022pearc,galleguillos2020,ozden2022elastisim,tang2009cluster} are primarily programming frameworks that demand simulation expertise to set up experiments and compare outcomes (\S\ref{sec:related}).
Support for online decision-making without future-workload knowledge is very limited. Moreover, the surrounding workflow, from ingesting scheduler records to configuring experiments and comparing outcomes, is rarely exposed through an interface designed for facility staff.
The resulting gap is not simply a lack of simulation methods, but a lack of operationally usable analytics that connects those methods to day-to-day facility decisions. Closing this gap is a central objective of this work.

We present \textbf{CQTwin}, an open-source toolkit that provides offline and online digital twins for HPC schedulers, together with an interactive dashboard, that turns operational scheduling data into actionable insights for facility staff, operators, and researchers.
Rather than building a simulator from scratch, CQTwin builds on CQSim~\cite{tang2009cluster,yang2013sc,fan2019hpdc,kurkure2025pads}, a Python-based, open-source discrete-event simulator refined through more than a decade of trace-driven scheduling research. CQTwin preserves CQSim's scheduling semantics, while adding a high-performance C++ discrete-event engine, an online event-streaming interface, and a browser-based environment for operational analysis.

\textbf{Offline twin: trace replay.}
In offline mode, CQTwin replays a complete production trace by submitting each job at its recorded arrival time and allowing a selected scheduling policy and backfilling strategy to determine when that job starts. This replay reveals how the same workload would have unfolded under a policy different from the one used in production. Users can compare per-job wait times, slowdown, system utilization, and fairness across policies without modifying or disrupting the production scheduler.

\textbf{Online twin: real-time decision support.}
In online mode, jobs enter the twin only as their arrival events are received. At each scheduling cycle, the twin therefore contains exactly the information that would be visible to a live scheduler. From this state, CQTwin can generate projections showing how each policy in a user-selected pool would process the current workload. It can also expose scheduling control to an external agent, enabling learning-based or adaptive policy selection~\cite{brockman2016gym}.

\textbf{Visual dashboard: analysis without scripting.}
CQTwin integrates these capabilities into a browser-based dashboard that supports the complete analysis workflow. Users can load a job trace in the Standard Workload Format (SWF)~\cite{SWF}, examine job sizes, runtimes, arrival patterns, and queue pressure, and then configure and launch simulations through a web form. Completed runs are summarized through interactive plots and tables, including wait-time distributions, utilization timelines, and side-by-side policy comparisons, without requiring post-processing scripts.
Moreover, the dashboard includes a \emph{Snapshot Lab} for state-level what-if analysis: users can freeze a simulated system state, inspect its queued and running jobs, and replay counterfactual interventions such as policy switches or planned system downtime.

Together, these components support an end-to-end operational workflow. For example, an operator observing rising wait times can load a recent trace, compare candidate scheduling policies, and inspect the resulting tradeoffs directly in the dashboard. The offline twin shows how each policy would have performed over the observed workload, while the online twin reveals, cycle by cycle and without hindsight, when an alternative policy would have processed the scheduler-visible workload more effectively. Analyses that previously required simulator expertise, custom scripts, and separate visualization pipelines can therefore be completed within a single interactive environment.

We validate CQTwin using production workload traces from the Argonne Leadership Computing Facility (ALCF) spanning 2016--2023. The C++ engine exactly reproduces Python CQSim's per-job wait times, with zero divergence across 50,000 jobs under five scheduling policies. This equivalence extends through the online path: a stream-fed online run exactly reproduces trace-driven replay across the complete 147,513-job Polaris-2023 trace, establishing an end-to-end fidelity chain from the event stream to the validated reference implementation. The engine replays a year-long production trace in 2.3~seconds, achieving up to a $24\times$ speedup over the original Python-based CQSim.
Per-cycle projections add approximately 0.66~ms of overhead while leaving the committed schedule unchanged, and a greedy controller demonstrates end-to-end external policy control by switching the committed policy 3,084 times during a year-long run through the Gym-style interface.

\section{Related Work}
\label{sec:related}

\textbf{Trace-driven scheduling simulation.}
Production workload traces allow researchers to hold the workload fixed while varying the scheduling policy. By replaying these traces through simulators, prior studies have evaluated FCFS and backfilling behavior~\cite{mu2001}, fault-aware utility scheduling~\cite{tang2009cluster}, energy-aware scheduling~\cite{yang2013sc}, multi-resource allocation~\cite{fan2019hpdc}, and reinforcement-learning-based dispatch~\cite{fan2021ipdps,zhang2020sc}.

Several open-source trace simulators have served as experimental platforms for such studies. Alea~\cite{klusacek2010alea}, built on the GridSim toolkit~\cite{buyya2002gridsim}, replays SWF workloads using built-in scheduling algorithms that users can extend at the source-code level. The Slurm Simulator~\cite{simakov2022pearc} takes a different approach from purpose-built simulators by adapting the production Slurm codebase itself, allowing offline studies of scheduler configurations to reproduce real Slurm behavior. AccaSim~\cite{galleguillos2020} provides a customizable event-driven framework for dispatching research, while ElastiSim~\cite{ozden2022elastisim} extends batch-system simulation to malleable jobs. These systems are capable experimental instruments, and some provide visualization of simulation outputs; Alea, for example, can display utilization and queue metrics during a run. Nevertheless, they are primarily operated as programming or configuration frameworks: setting up an experiment requires writing scheduler code or editing configuration files, while comparing policies across runs typically relies on user-developed analysis scripts. None packages the end-to-end workflow within a single interactive interface or supports evaluating alternative policies from an intermediate system state as a first-class capability.

\begin{figure*}[t]
\centering
\includegraphics[width=0.9\textwidth]{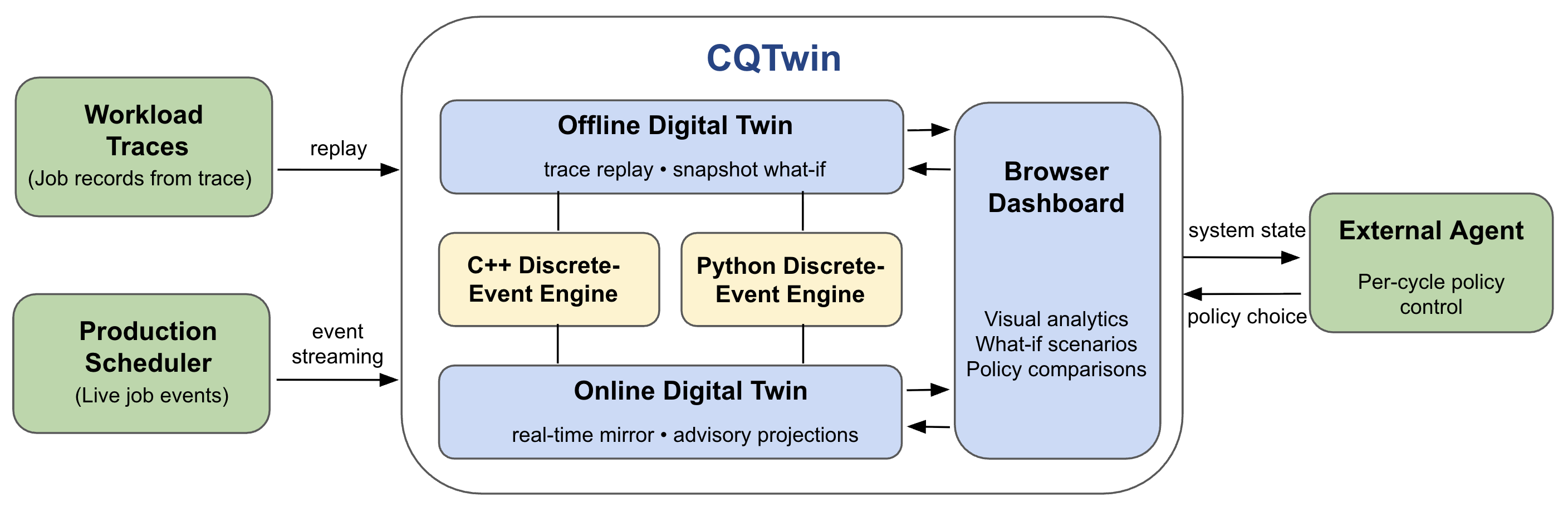}
\caption{CQTwin overview. Workload traces drive the offline twin, while live job events from a production scheduler drive the online twin through a streaming protocol and a site-specific bridge. Users configure simulations and inspect results through the browser dashboard, while an external agent can observe the system state and select the dispatch policy at each scheduling cycle.}
\label{fig:workflow}
\end{figure*}

\textbf{Digital twins of HPC systems.}
A complementary line of work develops digital twins for facility-level operational decision support. ExaDigiT~\cite{brewer2024sc} is an open framework for supercomputer digital twins that couples job-trace replay with resource-allocation, power, and cooling models. Its scheduling extension evaluates how policies and incentive structures affect facility-level outcomes~\cite{maiterth2025sc}. A digital twin of the AOBA system similarly reproduces job scheduling and power-management behavior to help administrators tune parameters before deployment~\cite{ohmura2022aoba}. These efforts demonstrate the operational value of twin-based what-if analysis~\cite{fan2021ipdps,zhang2020sc,maiterth2025sc}, but their fidelity depends on coupling scheduler behavior with telemetry and installation-specific power and cooling models, making them facility-scale infrastructure projects.

CQTwin addresses a lightweight, scheduler-focused portion of the same design space. It is portable across systems as long as scheduler records are represented in the SWF format, and runs locally without facility-level infrastructure integration. More importantly, CQTwin introduces an online digital twin capability absent from existing simulators, enabling real-time simulation from a live system state.

\textbf{Discrete-Event Scheduling Simulator CQSim.}
The Python-based CQSim has been developed and refined over more than a decade. It originated from Qsim, developed in 2009 as the simulation component of the Cobalt resource manager, the production batch scheduler at the Argonne Leadership Computing Facility. Qsim was released with Cobalt and replayed production Blue Gene workloads to support utility-based and fault-aware scheduling studies~\cite{tang2009cluster}. In 2011--2012, Qsim evolved into CQSim, a modular, open-source Python simulator supporting pluggable scheduling policies and SWF trace replay. CQSim subsequently served as the experimental platform for a wide range of studies, including energy-aware scheduling~\cite{yang2013sc}, multi-resource allocation~\cite{fan2019hpdc}, deep-reinforcement-learning-based scheduling~\cite{fan2021ipdps}, and CQSim+ for multi-system scheduling~\cite{kurkure2025pads}. CQTwin preserves CQSim's scheduling semantics, rooted in Qsim's origin as a component of the Cobalt scheduler~\cite{tang2009cluster} and validated through per-job agreement with the Python implementation (\S\ref{sec:eval}). It further extends CQSim with offline trace-replay and online event-driven simulation modes, together with an integrated browser-based dashboard designed for use by operators and researchers.

\section{CQTwin Design}
\label{sec:design}

\begin{table}[t]
\centering
\caption{Scheduling policies supported by CQTwin.}
\label{tab:policies}
\footnotesize
\setlength{\tabcolsep}{4pt}
\begin{tabular}{@{}ll@{}}
\toprule
\textbf{Policy} & \textbf{Priority criterion} \\
\midrule
FCFS, LCFS & Earliest / latest submit first \\
SJF, LJF   & Shortest / longest requested walltime \\
SRF, LRF   & Fewest / most requested processors \\
SCF, LCF   & Smallest / largest resource--time product \\
WFP1--WFP3 & Weighted wait-to-walltime (linear to cubic), size-scaled~\cite{allcock2017jsspp} \\
SWW        & Wait-to-walltime ratio weighted by job size \\
FCSJ       & Wait normalized by walltime and size \\
FAT        & Weighted wait scaled by fractional system size \\
UNICEP     & Wait normalized by processor fraction~\cite{allcock2017jsspp} \\
F1--F4     & Learned composite priority functions~\cite{carastan2017sc} \\
BSLD       & Bounded slowdown of the waiting job~\cite{mu2001} \\
\bottomrule
\end{tabular}
\end{table}

CQTwin is organized around a simple idea: a single, high-performance simulation core serves as a digital twin of the batch scheduler and can be accessed through the visual interface.
Figure~\ref{fig:workflow} shows how the components fit together. At the center is a C++ discrete-event engine, accompanied by the original Python-based CQSim engine~\cite{tang2009cluster,yang2013sc,fan2019hpdc}, which remains available to users who prefer to develop and extend scheduling policies in Python. The C++ engine is validated against Python CQSim and reproduces its scheduling decisions exactly on a per-job basis (\S\ref{sec:fidelity}); the two engines therefore preserve the same scheduling semantics.

The twin operates in two modes: offline, by replaying archived scheduler traces, and online, by consuming scheduler job events as they occur. Users can access both modes through a browser-based dashboard connected to a local server that configures runs, launches the engine, and transforms simulation outputs into interactive plots and tables. CQTwin also provides a programmatic interface through which external programs---such as reinforcement-learning agents or scripted experiments---can drive the online twin directly and select the dispatch policy at each scheduling cycle (\S\ref{sec:online}). The entire toolkit runs locally: it requires no cloud service or user account, and no operational data leaves the host machine. This is a deliberate design choice for facilities whose data-governance policies restrict movement of production job records.

\subsection{Discrete-Event Simulation Core}
\label{sec:core}

The engine simulates a batch-scheduled cluster as the real system experiences it: jobs arrive over time and wait in a queue, a scheduling policy orders the queued jobs, and each job starts when sufficient nodes become available and releases those nodes upon completion. Workloads are derived from scheduler records, SWF workload traces in offline mode and live scheduler event streams in online mode. The simulation distinguishes between the information available to the scheduler and the job's actual execution behavior: scheduling and backfilling decisions use each job's requested walltime, whereas resource occupancy is determined by its recorded runtime. An optional \emph{perfect} mode instead schedules jobs using their actual runtimes, providing an oracle upper bound on the potential benefit of perfectly accurate runtime estimates.

CQTwin provides twenty scheduling policies behind a common interface (Table~\ref{tab:policies}), from simple orderings to the weighted-priority functions used in production at ALCF, such as WFP and UNICEP~\cite{allcock2017jsspp}, and the machine-learned composite functions F1--F4 of Carastan-Santos and de Camargo~\cite{carastan2017sc}. All policies combine with three backfilling modes~\cite{lifka1995easy,mu2001}: none, EASY, and Conservative. Scheduled and unscheduled maintenance windows can be injected from a file, so policies can be evaluated against realistic operational disruptions.

The engine preserves CQSim's scheduling semantics: Section~\ref{sec:eval} verifies exact per-job agreement with the Python-based CQSim implementation. At the same time, it is fast enough to replay a year-long production trace in seconds. This performance makes the twin operationally useful: policy comparisons complete within an interactive time frame, and repeated simulations are inexpensive enough to run at every scheduling cycle, as required by the online twin.

\begin{figure*}[t]
\centering
\includegraphics[width=0.9\textwidth]{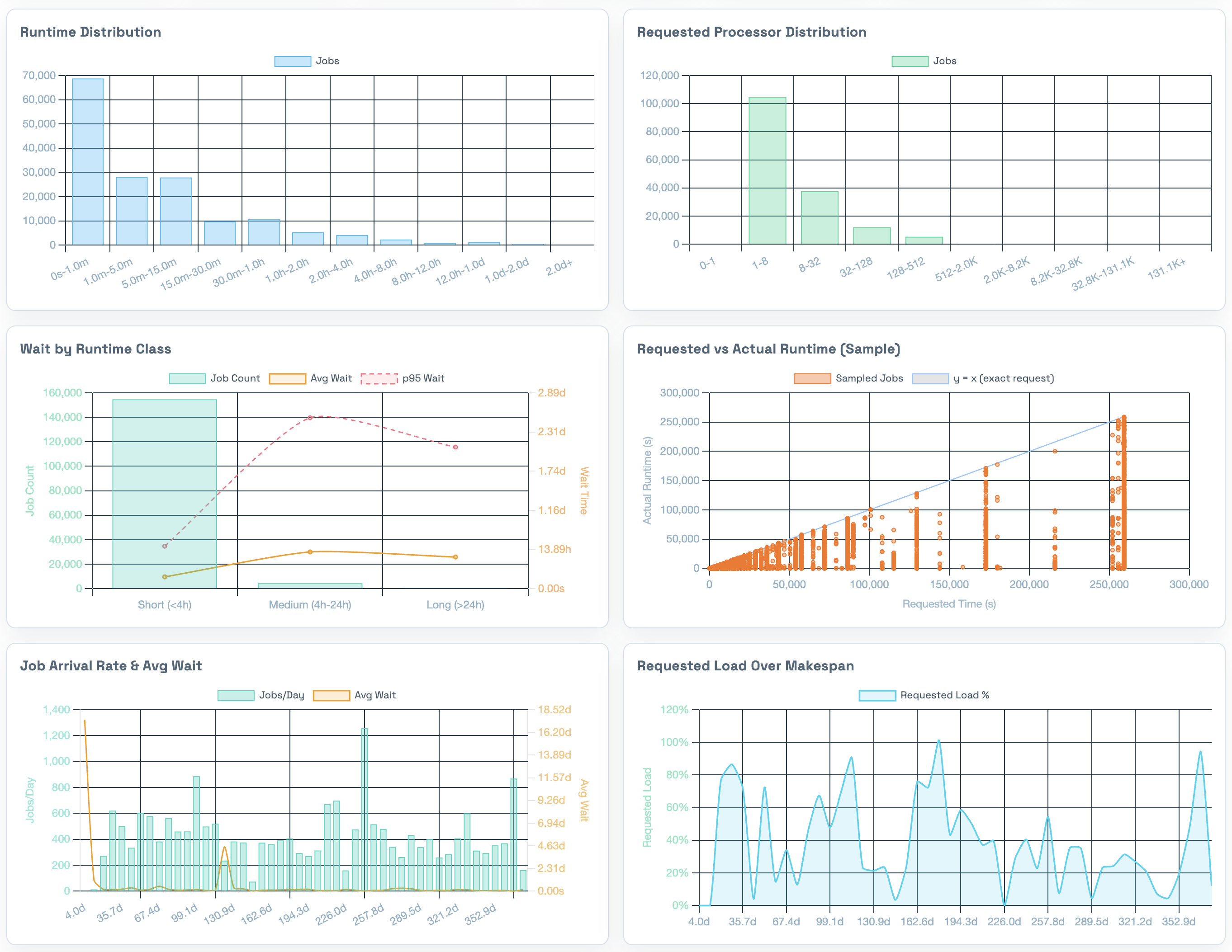}
\caption{Workload analysis view for the Polaris-2023 trace, computed from the SWF alone before any simulation. Requested load is reconstructed from recorded job records and can exceed 100\% when demand outstrips capacity.}
\label{fig:workload}
\end{figure*}

\subsection{Offline Digital Twin: Trace Replay and What-If Analysis}
\label{sec:offline}

\textbf{Trace replay.} The offline twin answers retrospective questions using a facility's accumulated scheduler records. The user selects a historical trace, a scheduling policy, and a backfilling mode; the engine then replays the complete workload under that configuration and reports how the system would have behaved, including each job's wait time and slowdown, system utilization over time, and fairness metrics. Because the same workload is used across runs, differences in outcomes can be attributed to the scheduling configurations under comparison. Multiple configurations can be launched together and compared side by side (Figure~\ref{fig:overlay}).
From an ODA perspective~\cite{netti2021oda}, the offline twin transforms archived scheduler logs into an analysis instrument. Scheduling policies can be evaluated against real production workloads without modifying the production system.

\textbf{Snapshot Lab: what-if analysis on frozen states.}
Beyond complete trace replay, the offline twin supports exploratory what-if analysis through the Snapshot Lab, which operates on completed simulation runs. A user can freeze a run at any selected time and inspect the system as it appeared at that moment, including queued jobs, running jobs, available nodes, and the next expected completion event.
Selecting a queued job produces a plain-language explanation of why it is waiting. The explanation includes its queue position and dispatch rank, the running jobs occupying the resources it requires and their expected completion times, its earliest capacity-constrained start time, and an estimated wait time with a confidence indicator derived from similar completed jobs.

For example, freezing a run during a burst of job arrivals captures the system when its queue is deepest. The explanation view identifies which running jobs hold the nodes required by the head-of-queue request and when those nodes will become available, while a forked scenario shows how the same backlog would drain under an alternative scheduling policy. From the frozen state, the user can create what-if scenarios and compare them with the baseline: switching the scheduling policy to examine changes in projected wait times, pinning a job as the next dispatch candidate, or injecting a planned maintenance window to measure queue-level delays. A downtime advisor can rank candidate maintenance windows by projected impact. Scenarios can be saved, reloaded, and viewed side by side.

\subsection{Online Digital Twin: Real-Time Decision Support}
\label{sec:online}

The online twin addresses a different question: not \emph{what would have happened over the complete trace}, but \emph{what is the appropriate decision now}. The twin is event-driven: jobs enter only when their submission events are received, so its state at each scheduling cycle contains the current queue and running jobs but no information about future submissions.

Events reach the twin in one of two ways. First, a recorded trace can drive the twin directly, with jobs admitted strictly at their recorded submission times. Second, the twin can consume a \emph{job-event stream} through its socket protocol. An external feeder submits jobs as they arrive and indicates the time through which the stream is complete. The twin refuses to simulate beyond this asserted frontier: if the next simulation event lies beyond it, the twin waits for additional input rather than assuming that no jobs will arrive. The no-future-knowledge property is therefore enforced by construction: the twin process never contains a job event that it has not yet received. CQTwin includes a feeder that replays recorded traces as timed streams; for live schedulers, it documents an adapter interface specifying how a site-specific bridge would map scheduler events into the same protocol.

CQTwin supports real-time decision-making. At each scheduling cycle, the live state can be forked across a user-selected pool of policies, with each fork simulated to estimate how the corresponding policy would process the currently visible workload without altering the committed schedule. Alternatively, an external program can interface with the production scheduler, observe the current state and optional projections, and select the dispatch policy at each cycle. Together, these capabilities make CQTwin a training and evaluation environment for learning-based and adaptive scheduling~\cite{fan2021ipdps,kurkure2026mars,maiterth2025sc}.

\subsection{Visual Dashboard: Scheduler-Data Analytics}
\label{sec:dashboard}

The dashboard is a browser-based visual analytics interface backed by a local server; no command-line interaction is required at any step. It serves both twins, and analysis proceeds in three stages: before, during, and after simulation.

\textbf{Before simulation}, the workload analysis view characterizes a loaded trace from the SWF alone---no simulation required---across six panels: runtime and requested-processor distributions, wait time by runtime class, requested versus actual runtime, job arrival rate against average wait, and requested load over the makespan (Figure~\ref{fig:workload}). This triage layer is operational data analytics in its own right, and often answers questions by itself. On Polaris-2023, for example, it immediately reveals a workload dominated by short, small jobs---91\% finish within an hour, and 69\% of submissions request eight or fewer nodes yet account for under 6\% of requested processor-hours, while the top 5\% of jobs by size account for 65\%---alongside heavily overestimated walltimes (jobs run to only 17\% of their request on average, the slack that backfilling exploits) and bursty queue pressure (arrivals average 420 jobs per day but spike past 1,250). All of this is visible at trace-load time and guides which policies are worth simulating.

\begin{figure*}[t]
\centering
\includegraphics[width=0.9\textwidth]{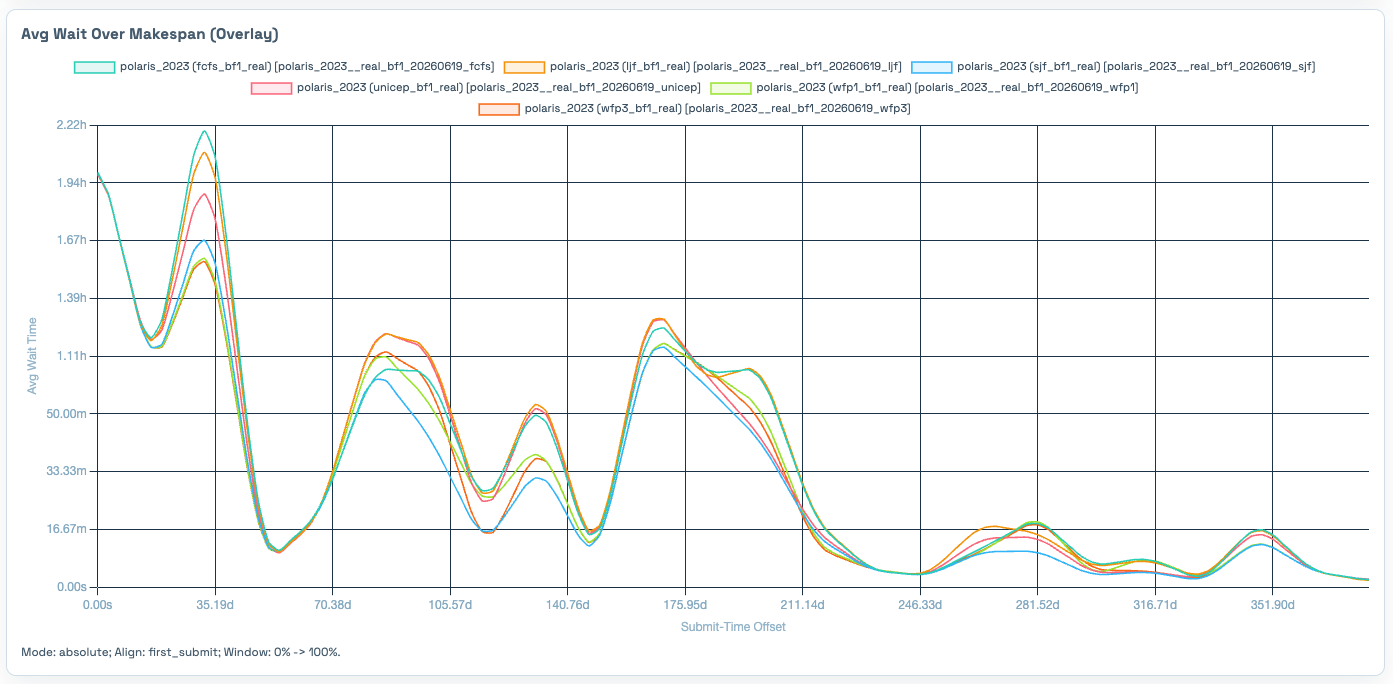}
\caption{Dashboard comparison overlay: average wait within submission-time bins over the makespan, six policies on full Polaris-2023 (EASY backfill).}
\label{fig:overlay}
\end{figure*}

\textbf{During simulation}, a setup form organizes configuration into sections for trace selection, scheduling policy, backfilling and window options, time range, and cluster scaling, plus a compatibility section that accepts legacy CQSim configuration fields directly; a parameter sweep section launches a full policy-backfill-window cross-product in one submission, and progress streams live to the browser. An online projection run is configured the same way, adding the policy pool and the projection horizon and interval.

\textbf{After simulation}, the results view presents completed runs together, shows deltas against a chosen baseline, and overlays wait time dynamics across the makespan; Figure~\ref{fig:overlay} shows this comparison for six policies on Polaris-2023. The question ``which policy, and when does it matter?'' is answered visually rather than by scripting over CSV files---turning raw simulation output into the kind of sharable, at-a-glance evidence that operations meetings and procurement discussions actually use.

\section{Experiments}
\label{sec:eval}

We evaluate CQTwin along three dimensions: the fidelity of its offline and online twins, simulation performance, and online-twin overhead. CQTwin is compiled with \texttt{-O3}, while Python CQSim runs under CPython~3.9. All reported timing results are the medians of three runs. Table~\ref{tab:traces} summarizes the three ALCF production traces used in our evaluation, spanning nearly a decade and three system architectures: the IBM Blue Gene/Q Mira, the Cray Theta, and the GPU-accelerated Polaris. The 2023 Polaris trace covers the system's first full year of production operation. All three traces are publicly available through the ALCF data reports~\cite{alcfdata}.

\begin{table}[t]
\centering
\caption{ALCF production traces and simulation wall-clock time in seconds: CQTwin vs.\ Python CQSim. FCFS + EASY backfill, single thread, median of three runs.}
\label{tab:traces}
\small
\setlength{\tabcolsep}{3.5pt}
\begin{tabular}{lrrrrc}
\toprule
\textbf{Trace} & \textbf{Nodes} & \textbf{Jobs} & \textbf{Python} & \textbf{C++} & \textbf{Speedup} \\
\midrule
Mira-2016    & 49,152 &   50,611 & 19.29 & 0.79 & $24.4\times$ \\
Theta-2021   &  4,360 &   63,336 & 16.32 & 0.85 & $19.2\times$ \\
Polaris-2023 &    560 & 147,513 & 55.80 & 2.29 & $24.4\times$ \\
\bottomrule
\end{tabular}
\end{table}

\subsection{Simulation Fidelity}
\label{sec:fidelity}

\textbf{Offline fidelity.}\; We verify that the C++ engine reproduces Python CQSim's scheduling decisions by simulating the first 50,000 jobs of Theta-2021 under EASY backfill across five policies (FCFS, SJF, LJF, WFP1, WFP3) and comparing per-job wait times. Every one of the 50,000 jobs matches the Python CQSim wait time exactly under every policy---zero mean absolute error and 100\% exact agreement. The exact match is stronger evidence than it may first appear. WFP1 and WFP3 compute floating-point priority scores, so two implementations could plausibly diverge in rounding or in how they order jobs with equal priority---yet even a single such divergence would have shifted some job's start time and produced a non-zero wait-time error. Zero error across all 50,000 jobs therefore means the C++ engine reproduces not just the policy formulas but Python CQSim's tie-breaking and dispatch order exactly.

\begin{figure*}[t]
\centering
\includegraphics[width=0.9\textwidth]{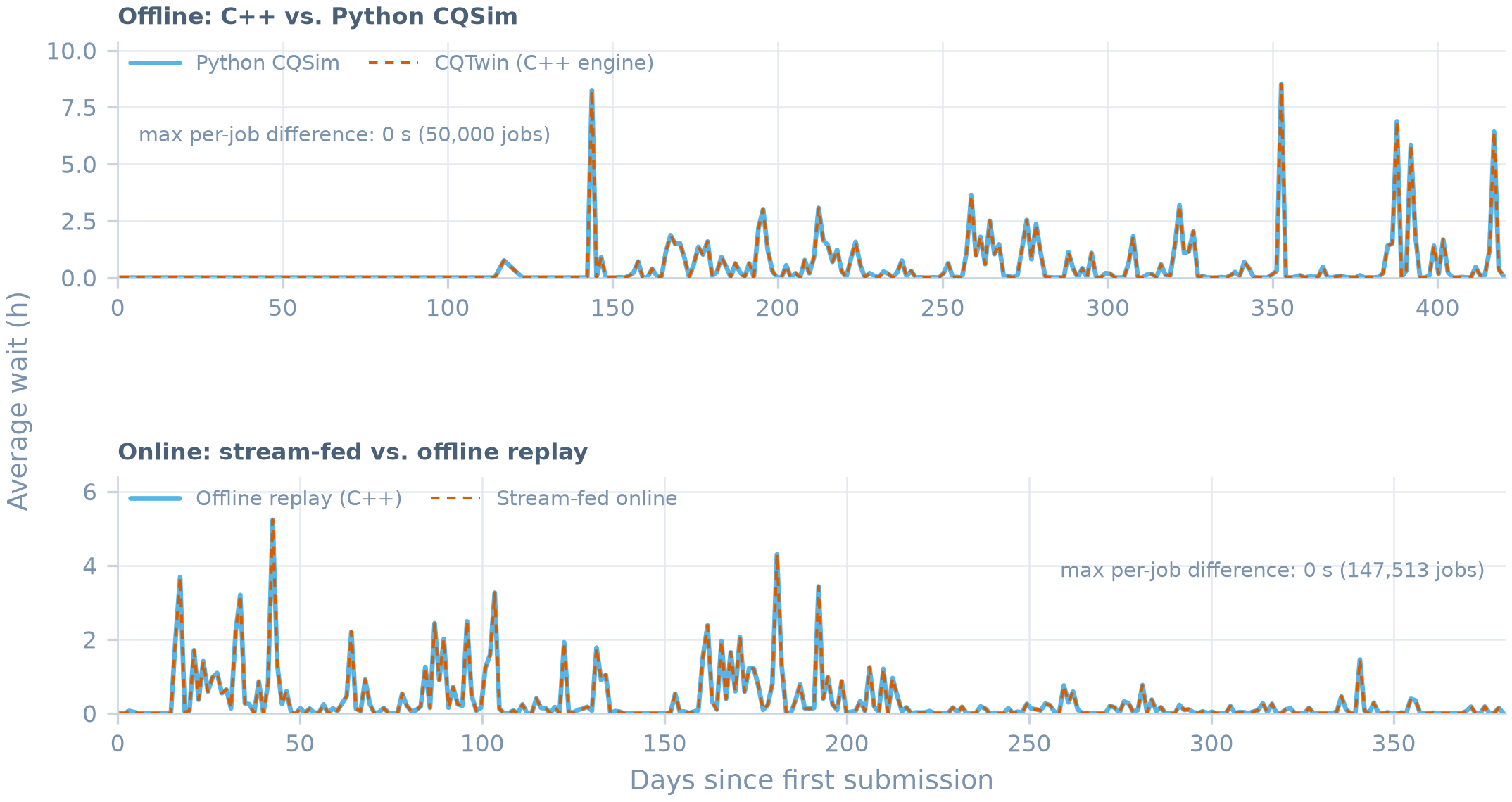}
\caption{Fidelity over the makespan: average wait within submission-time bins; in both panels the two curves coincide at every bin, reflecting exact per-job agreement. Top (offline): Python CQSim vs.\ the C++ engine on the first 50,000 jobs of Theta-2021 (WFP3, EASY backfill). Bottom (online): offline replay vs.\ the stream-fed online twin on full Polaris-2023 (WFP3 committed, EASY), submission times origin-normalized.}
\label{fig:fidelity}
\end{figure*}

\textbf{Online fidelity.}\; The online twin is validated against four properties, each enforced by the automated test suite. First, \emph{fidelity}: an online run with a fixed per-cycle policy choice reproduces the offline engine's per-job schedules exactly---wait times, and start times up to the documented time origin---with zero divergent jobs across EASY and no-backfill configurations. Second, \emph{stream equivalence}: driving the twin from a job-event stream reproduces the trace-driven results exactly, per job, across the full 147,513-job Polaris-2023 trace, in both bulk and per-event-group pacing and with projections enabled. Combined with the offline result above, this closes the chain (Figure~\ref{fig:fidelity}): stream-fed online equals trace-fed online equals the offline engine, which equals Python CQSim. Third, \emph{no future visibility}: projection forks are verified to contain no job whose submit time exceeds the current simulation time; in stream mode this is also structural, since the twin never holds unsent jobs. Fourth, \emph{non-interference}: the committed schedule is identical whether projections are disabled or enabled at any interval---the full-trace average wait agrees to twelve significant figures across all projection intervals tested.

\subsection{Simulation Speed}

As Table~\ref{tab:traces} shows, CQTwin achieves up to $24\times$ speedup over Python CQSim under FCFS with EASY backfill, completing every trace in under 2.5~seconds; the year-long Polaris trace simulates in 2.3~seconds. The speedup comes from moving the event loop, job state, and policy dispatch into compiled code with incremental event submission and lean per-run state. In practical terms, a six-policy comparison completes in about a minute through the dashboard, and workflows built on repeated simulation---parameter sweeps, per-cycle projections, and training loops that invoke the simulator thousands of times---drop from days of compute to hours.

\begin{table}[t]
\centering
\caption{Online overhead, full Polaris-2023 (WFP3 committed, three-policy pool, EASY): end-to-end wall-clock time for the full trace and per-cycle projection cost. Average wait is identical in all configurations.}
\label{tab:overhead}
\small
\begin{tabular}{lrrr}
\toprule
\textbf{Configuration} & \textbf{Wall (s)} & \textbf{ms/cycle} & \textbf{Avg wait (h)} \\
\midrule
Projections off           &  26.4 &  --- & 0.384 \\
Projections, interval 1   & 164.6 & 0.66 & 0.384 \\
Projections, interval 10  &  41.1 & 0.07 & 0.384 \\
Projections, interval 100 &  28.2 & 0.01 & 0.384 \\
\bottomrule
\end{tabular}
\end{table}

\subsection{Online Overhead}
\label{sec:onlineeval}

Per-cycle projections are cheap. On the full Polaris-2023 trace, forking and replaying all three pool policies costs about 0.66~ms per cycle, and even projecting at every cycle completes the year-long trace in under three minutes (Table~\ref{tab:overhead}). The overhead buys information without side effects: the committed average wait is identical whether projections run at every cycle or not at all. The same projections drive external policy control end to end---a greedy controller ran the full trace through the Gym-style interface, switching the committed policy 3,084 times.
This demonstrates that projection outputs can drive an external control loop through a standard interface. We deliberately do not evaluate whether greedy switching improves scheduling outcomes, as CQTwin's goal is operational analytics and decision support rather than developing a new scheduling method.

\section{Discussion}
\label{sec:discussion}

CQTwin replays scheduler records, making it portable across workload traces but limiting the behaviors it can represent. SWF-style records do not capture job dependencies, advance reservations, queue-specific access rules, topology constraints, or multi-resource requirements such as GPUs and memory. CQTwin is therefore a scheduler-focused digital twin rather than a full production-scheduler emulator. Exact per-job agreement with Python CQSim is validated for five policies under EASY backfilling, and the online twin inherits this fidelity through fixed-policy and stream-equivalence tests.

The online twin currently operates on recorded scheduler events rather than a validated live production feed. Its online property is therefore about \emph{information availability}:
jobs become visible only when they arrive, and each projection is computed from the state available at that decision point, without access to future submissions.
The validated event-stream interface reproduces trace-driven results exactly, so production deployment requires only a site-specific bridge that maps scheduler events into an event stream, without modifying the CQTwin engine.
Future arrivals are unknowable at decision time, so a projection is a reference outcome rather than a prediction: it answers how a scheduling policy would clear the queue from the current state. CQTwin is an analytics tool, not a scheduling method: the committed policy always drives the actual run, while projections provide evidence for analysis or external policy control.
Overall, CQTwin provides a fast, high-fidelity, and visualization-friendly foundation for HPC batch-scheduling research in both offline and online settings. Its modular interfaces also provide a flexible basis for developing and evaluating more advanced scheduling methods, allowing future intelligent and adaptive scheduling techniques to be built on top of CQTwin.

Currently CQTwin models whole-node allocation only and does not represent GPUs, memory, heterogeneous node types, or other multi-resource constraints. It also excludes energy, power, cooling, temperature, and other facility telemetry modeled by broader digital twins such as ExaDigiT~\cite{brewer2024sc}.

\section{Conclusion}
\label{sec:conclusion}

HPC facilities collect scheduler data, but converting these records into practical insight remains difficult. CQTwin addresses this gap by providing an open-source, scheduler-focused digital twin that supports both offline trace replay and online decision guidance without future-workload knowledge. Built on the CQSim simulator family, CQTwin combines a fidelity-validated, high-performance C++ engine with an interactive browser-based dashboard and Snapshot Lab, enabling facility staff, operators, and researchers to analyze workloads, compare scheduling policies, and explore what-if scenarios without modifying the production system or transferring operational data off site.
Our evaluation shows that the C++ engine reproduces Python CQSim's scheduling decisions exactly on a per-job basis while running up to $24\times$ faster; stream-fed online execution exactly reproduces trace-driven replay; per-cycle online simulation adds approximately 0.66~ms of overhead per decision cycle; and end-to-end external policy control is demonstrated through the Gym-style interface.

\section*{Code and Data Availability}
CQTwin is available at
\url{https://github.com/SPEAR-UIC/CQTwin}. The ALCF production
traces used in our evaluation (Mira-2016, Theta-2021, and
Polaris-2023) are publicly available through the ALCF data
reports at \url{https://reports.alcf.anl.gov/data/index.html}.

\section*{Acknowledgment}
This work was supported in part by the U.S. National Science Foundation under Grants OAC-2402901 and CCF-2515009 and by the U.S. Department of Energy under Contract No. DE-SC0024271.
\bibliographystyle{IEEEtran}
\bibliography{references}

\end{document}